\documentclass[useAMS,usenatbib]{mn2e}
\usepackage{aas_macros,graphicx,times,multirow,amsmath}
\usepackage[]{color,hyperref}

\title[X-ray observations of sdB binaries with compact companions]{Constraints on the   winds of hot subdwarf stars from X-ray observations of two sdB binaries with compact companions: \\
 \cd\ and \pg\   
   }
\author[S. Mereghetti et al.]{S.~Mereghetti,$^{1}$\thanks{E-mail: \href{mailto:sandro@iasf-milano.inaf.it}{sandro@iasf-milano.inaf.it}}
N.~La Palombara,$^{1}$ P.~Esposito,$^{1}$ F.~Gastaldello,$^{1,2}$ A.~Tiengo,$^{1,3,4}$
\newauthor U.~Heber,$^{5}$ S. Geier$^{5,6}$, J. Wilms  $^5$ 
\smallskip\\
$^1$INAF -- Istituto di Astrofisica Spaziale e Fisica Cosmica - Milano, via E. Bassini 15, I-20133 Milano, Italy\\
$^2$Department of Physics and Astronomy, University of California at Irvine, 4129, Frederick Reines Hall, Irvine, CA 926
97-4575\\
$^3$IUSS -- Istituto Universitario di Studi Superiori, piazza della Vittoria 15, I-27100 Pavia, Italy\\
$^4$INFN -- Istituto Nazionale di Fisica Nucleare - Pavia, via Bassi 6, I-27100 Pavia, Italy\\
$^5$Dr. Karl Remeis-Observatory \& ECAP, Friedrich-Alexander University Erlangen-Nuremberg, Sternwartstr. 7,
 D-96049 Bamberg, Germany\\
 $^6$European Southern Observatory, Karl-Schwarzschild-Str. 2, 85748 Garching, Germany
}
\date{Accepted 2014 April 16. Received 2014 April 16; in
original form 2014 March 21} \pagerange{\pageref{firstpage}--\pageref{lastpage}} \pubyear{2014}

\def\LaTeX{L\kern-.36em\raise.3ex\hbox{a}\kern-.15em
    T\kern-.1667em\lower.7ex\hbox{E}\kern-.125emX}

\def\xmm {\emph{XMM-Newton}}

\def\sw {\emph{Swift}}

\def\hd {HD\,49798}
\def\bd {BD\, +37$^\circ$\,442}
\def\cd {CD\, --30$^\circ$\,11223}
\def\pg {PG 1232--136}

\def\msun{~\rm{M}_{\odot}}
\def\rsun{~R_{\odot}}

\def\mdot {\dot M}

\def \arcmin {\hbox{$^\prime$}}
\def \arcsec {\hbox{$^{\prime\prime}$}}

\def\spose#1{\hbox to 0pt{#1\hss}}
\def\ltsim{$\mathrel{\spose{\lower 3pt\hbox{$\sim$}}
       \raise 2.0pt\hbox{$<$}}$\thinspace}
\def\gtsim{$\mathrel{\spose{\lower 3pt\hbox{$\sim$}}
       \raise 2.0pt\hbox{$>$}}$\thinspace}
\newcommand\solar{\hbox{{$Z_{\odot}$}}}

\newcommand{\source}{\mbox{XMMU\,J141057.7-305132}}
\newcommand{\apec}{APEC}

\newcommand{\fxunits}{\mbox{erg cm$^{-2}$ s$^{-1}$}}
\newcommand{\lxunits}{\mbox{erg s$^{-1}$}}
\newcommand{\lx }{${L_{\mathrm{X}}}$}
\newcommand{\tx }{${T_{\mathrm{X}}}$}
\newcommand{\ned}{{\em{NED}}}
\newcommand\omegam{\hbox{{$\Omega_{\rm m}$}}}
\newcommand\omegalambda{\hbox{{$\Omega_{\Lambda}$}}}
\newcommand\kmsmpc{{\rm km s$^{-1}$ Mpc$^{-1}$}}
\newcommand\ho{\hbox{{$H_{0}$}}}

\begin{document}

\label{firstpage}
\maketitle
\begin{abstract}
Little observational data  are  available on the weak stellar winds of hot subdwarf stars of B spectral type (sdB). Close binary systems composed of an sdB star and a compact object (white dwarf, neutron star or black hole) could be detected as accretion-powered X-ray sources. The study of their X-ray emission can probe the properties of line-driven winds of sdB stars that can not be derived directly from  spectroscopy because of the low luminosity of these stars. Here we report on the first sensitive X-ray observations of two sdB binaries with compact companions.  \cd\ is the sdB binary with the shortest known orbital period (1.2~h) and its companion is certainly a white dwarf.    \pg\ is  an  sdB binary considered the best candidate to host a black hole companion. We observed these stars with \xmm\ in August 2013 for 50~ks and in July 2009 for 36~ks, respectively.
None of them was detected and we derived luminosity upper limits of $\sim1.5\times10^{29}$~erg~s$^{-1}$ for \cd\ 
and $\sim5\times10^{29}$~erg~s$^{-1}$ for \pg . The corresponding mass loss rate for \pg\ is poorly constrained, owing to the unknown efficiency for black hole accretion. On the other hand, in the case of \cd\ we could derive, under reasonable assumptions, an upper limit of $\sim3\times10^{-13}$ $\msun~yr^{-1}$ on the wind mass loss rate from the sdB star. This is one of the few observational constraints on the weak winds expected in this class of low mass hot stars. 
We also report the results on the X-ray emission from a cluster of galaxies serendipitously discovered in the field of \cd .
\end{abstract}


\section{Introduction}

Luminous hot  stars are characterized by strong  winds  with mass loss rates $\mdot_{\mathrm{W}}\sim10^{-7}$--$10^{-5}$~$\msun$~yr$^{-1}$  and terminal velocities reaching a few thousands km s$^{-1}$ (see, e.g., \citealt{kud00}).  
These winds are driven  by  the resonant absorption and re-emission of the star optical/UV photons  by the wind atoms, which results in a net gain of radial momentum.  
The theory of line-driven winds predicts that the mass-loss rate scale with the luminosity approximately as $L^{1.5}$ and depend on the wind composition, since most of the relevant spectral lines are provided by metals. 
The observations of  O- and B-type stars of different luminosity, from main sequence to supergiants, agree reasonably well with these general predictions  \citep{pul08}. 

Much less is known, from the observational point of view, about the winds of less luminous hot stars, which, according to the models, should have weaker winds. 
Evidence of mass loss  has been seen in the central stars of planetary nebulae, as well as  in a few extreme helium stars and O-type subdwarfs,  with  $\mdot_{\mathrm{W}}$  as low as $\sim$$10^{-10}  \msun$~yr$^{-1}$ \citep{ham10}. 
It is not clear if the theory and scaling relations derived for luminous stars can be simply extended to such weak winds. 

Hot subdwarfs are evolved low--mass stars that have lost most of their hydrogen envelopes and are now in the stage of helium core burning (see \citealt{heb09} for a review).  
They are spectroscopically classified in  the mostly hydrogen-rich sdB (with effective temperature $T\sim25,000$--40,000 K) and the predominantly helium-rich  sdO stars (with $T$ \gtsim 40,000 K).
The abundance patterns of sdBs are strongly affected by atomic diffusion processes, that is by gravitational settling and radiative levitation leading to the depletion of some elements, e.g. helium, and strong enhancement of heavy metals by factors of up to 10$^4$ \citep{oto06,bla08,nas11}.
The  possible presence of winds in hot subdwarfs has been invoked to explain some of the abundance anomalies observed in these stars. Early diffusion modelling  \citep{mic89} predicted that the helium abundances should be a hundred times lower than the average observed ones. 
Since the time scale for helium diffusion ($\approx$10$^4$ yr) is much shorter than the extreme horizontal branch (EHB) life time ($\approx$10$^8$ yr), equilibrium abundances should be established rapidly.
To slow down helium diffusion, a stellar wind has been suggested \citep{fon97}. Indeed \citet{ung01} showed that the observed helium abundances can be explained if a stellar wind with $10^{-14}  \msun$~yr$^{-1}<\mdot_{\mathrm{W}}<10^{-12}  \msun$~yr$^{-1}$ is present.
\citet{vin02} calculated mass loss rates for EHB stars based on the line-driven wind theory and give upper limits for the mass loss rates of  $\mdot_{\mathrm{W}}<10^{-11}  \msun$~yr$^{-1}$ . For weak mass loss rates ($\mdot_{\mathrm{W}}<10^{-12}  \msun$~yr$^{-1}$ ), calculations by  \citet{ung08,ung08b}  show that the wind might fractionate, that is metals decouple from hydrogen and helium, and for rates below $\mdot_{\mathrm{W}}<10^{-16}  \msun$~yr$^{-1}$ become purely metallic. Since the wind is driven by metal lines, the results depend on the adopted metallicity or more precisely on the abundance of those ions that contribute most to the radiation pressure. Selective winds could explain some anomalous metal abundances, but fail to explain the observed helium abundances. An alternative explanation was offered by  \citet{mic11}, who considered turbulent mixing of the outer 10$^{-7} \msun$ to reproduce the observed abundance anomalies. 

No observational evidence for a stellar wind in a sdB star has yet been found. In optical spectra, the H$\alpha$ line profile is the most sensitive indicator for the presence of a stellar wind. 
\citet{heb03}  analysed the  H$\alpha$ line profiles of a sample of sdB and sdOB stars using the hydrostatic, fully metal line blanketed LTE model atmospheres. 
For all sdB stars the  H$\alpha$  profiles were matched indicating no evidence for a stellar wind. Only in the case of two sdOB stars the lines were not matched and slight asymmetries in the profile may hint at the presence of a stellar wind. Those objects, however, have already evolved off the EHB and are considerably more luminous than the sdB stars under study in this paper.

In view of the conflicting results of diffusion modelling it is of utmost importance to detect signatures of stellar winds observationally.
Such an opportunity arises if the sdB is orbited by a compact companion.
About two thirds of the sdB stars  are in close binary systems \citep{max01}, supporting the idea that  non-conservative mass transfer during a common envelope phase caused the loss of the massive hydrogen envelopes necessary to form hot subdwarfs.
Both   evolutionary computations and radial velocity surveys indicate that sdB binaries  contain compact objects, i.e. white dwarfs (WD) or, possibly but less frequently neutron stars (NS) or black holes \citep{han02,gei11}.
These systems can be detected as accretion-powered  X-ray sources, if some of the mass lost from the subdwarf  is transferred  at a sufficiently high rate onto the compact object. Their study can thus provide a way to investigate the properties of the sdB winds, as is routinely done for more luminous O and B stars.

Up to now, the only hot subdwarfs detected in the   X-ray range are two sdO stars: \hd\ and \bd .
\hd\  is an extensively studied single-lined spectroscopic binary in a 1.5 days orbit with a 13.2 s X-ray pulsar, most likely a massive  WD \citep{mer09}.   
\bd\ has similar optical properties, but it was believed to be a single star until the recent discovery
of X-ray emission with a periodic modulation at  19.2 s \citep{lap12},  suggesting also in this case the presence of a compact companion.  
Compared to the majority of hot subdwarfs, \hd\ and \bd\ have relatively strong winds  with   $\mdot_{\mathrm{W}}\sim3\times10^{-9} \msun$~yr$^{-1}$ \citep{ham10}.  Their X-ray luminosity  is consistent with the accretion rate expected from their wind properties.

A search for X-ray emission from a sample of 
candidate sdB+WD/NS binaries was carried out  with the \sw\ satellite, but none of the targets  was detected. The derived upper limits on their X-ray luminosity,
of the order of $L_{\mathrm{X}}\sim10^{30}$--$10^{31}$ erg s$^{-1}$ \citep{mer11b},  indicate sdB mass loss rates  $\mdot_{\mathrm{W}}<10^{-13}$--$10^{-12} \msun$~yr$^{-1}$, in the hypothesis of NS companions. 
On the other hand, if  the compact objects in these sdB binaries are WD, the implied limits on $\mdot_{\mathrm{W}}$  are about three orders of magnitude higher and not particularly
constraining.

Here we report on  sensitive X-ray observations  of  two particularly interesting sdB  binaries carried out with    the \xmm\ satellite. Our targets are \cd ,  which is the  tightest known sdB+WD binary, 
and \pg  , which is the best candidate sdB binary with a black hole companion. We also report the results on the X-ray emission from a cluster of galaxies serendipitously discovered in the field of \cd .

\begin{table*}
\caption{Parameters of the target sdB binaries.}
\label{parameters}

\begin{tabular}{lccclcl}

\hline

Parameter               &   Units      &  \multicolumn{2}{c}{\cd\ }    & References          &  \pg\  & References  \\
                                &              &  Solution 1     & Solution 2     &   &   &    \\
\hline

Orbital period                  &  [d]                & \multicolumn{2}{c}{0.048979072$\pm$0.000000002}   & 1   &  0.3630$\pm$0.0003  &   2 \\
Effective temperature       & [K]                 & \multicolumn{2}{c}{ 29200$\pm$400}                            & 1          & 26900$\pm$500  & 3 \\
Surface gravity (log g)    & [cm s$^{-2}$]  & \multicolumn{2}{c}{ 5.66$\pm$0.05}                              & 1         & 5.71$\pm$0.05  &  3\\
Helium abundance  (log y) &                    & \multicolumn{2}{c}{ --1.50$\pm$0.07}                           & 1          &  --1.47  & 4 \\
sdB mass                          &  [$\msun$]   &  0.47$\pm$0.03          &   0.54$\pm$0.02               & 1    &   0.45$^a$ & \\
sdB radius                       &    [$\rsun$]    &  0.169$\pm$0.005     & 0.179$\pm$0.003              & 1   & 0.16$\pm$0.01 &  3\\
Companion mass            & [$\msun$]     & 0.74$\pm$0.02            &  0.79$\pm$0.01                & 1      &   $>$6   &  3\\
Companion radius          & [$\rsun$]       & 0.0100$\pm$0.0004   &  0.0106$\pm$0.0002         & 1     &  --  & \\
Orbital inclination           &  [deg]            & 83.8$\pm$0.6           &  82.9$\pm$0.4                     & 1     &  $<$14 &  3 \\
Orbital separation          & [$\rsun$]       & 0.599$\pm$0.009        &  0.619$\pm$0.005            & 1        &   $>$ 4  & 3\\
Roche-lobe radius         & [$\rsun$]        &  0.204                        &      0.214                              &             &        0.75 &   \\
Distance                        &    [pc]             & \multicolumn{2}{c}{364$\pm$31}                               & 1                             & 570& 5 \\
X-ray luminosity           & [erg s$^{-1}$] &    \multicolumn{2}{c}{$<1.5\times10^{29}$ }               &  6 & $<5\times10^{29}$   & 6 \\
\hline
\end{tabular}

References:    1) \citet{gei13},   2)  \citet{ede05}, 3)  \citet{gei10}, 4)    \citet{saf94}, 5) \citet{alt04}, 6) this work 
 
$^{a)}$ Assumed value.
\end{table*}

\section{The targets}

\subsection{ \cd\  }
 
The V=12.3  magnitude star \cd , spectroscopically classified as  an  sdB  \citep{ven11},   has been included among the targets of the  MUCHFUSS project, which aims at finding hot subdwarf binaries with massive companions through optical spectroscopy and photometry \citep{gei11}.
Its orbital period of 1.2 hours, by far the shortest of any sdB binary,  was independently discovered by two groups \citep{ven12,gei13}.
The optical light curve shows eclipses and variations due to the ellipsoidal deformation of the sdB star caused by the tidal drag of its compact companion. Detailed modeling of these variations and extensive time-series spectroscopy led to an
accurate determination of the orbital parameters and of the properties of the sdB star. 
Assuming that the sdB rotation is tidally locked with the orbital period, as expected for such a tight system, gives   $M_{\mathrm{sdB}}= 0.47\pm0.03$ $\msun$    and $M_{\mathrm{WD}} = 0.74\pm0.02$ $\msun$ for the sdB and the WD mass, respectively.   
Relaxing the corotation constraint  leads to slightly different values 
of  $M_{\mathrm{sdB}} = 0.54\pm0.02$ $\msun$  and $M_{\mathrm{WD}} = 0.79\pm0.02$ $\msun$ (see Table \ref{parameters}  for the other parameters corresponding to these two solutions).

The very short orbital period and the certain presence of a WD companion make this system a promising target for X-ray observations. 
\cd\ is also interesting for its future evolution. 
\citet{gei13} proposed that it might be the progenitor of a type Ia supernova via the so-called sub-Chandrasekhar double-detonation scenario,  according to which carbon burning in the core can be triggered by the ignition of helium on the surface of an accreting WD \citep{fin10}. 
             
\subsection{ \pg\ }

\pg\ is a single-lined spectroscopic binary with orbital period of 0.36 days and  mass function of 0.0819 $\msun$ \citep{ede05}.  The   upper limit of 5 km s$^{-1}$ on its projected rotational velocity, implies, with the reasonable assumption of tidally locked orbital synchronisation, a system inclination $i<14^{\circ}$, hence a minimum companion mass of 6 $\msun$ \citep{gei10}.  It is therefore very likely that the companion star of \pg\ is a black hole, since any non-degenerate star of such a high mass would clearly appear in the optical spectra. 

A  short  X-ray observation of \pg\    with the \emph{Swift}/XRT instrument gave a 3$\sigma$ upper limit of   $1.8\times10^{-3}$ counts s$^{-1}$ on its 0.3--10 keV count rate, which  corresponds to a luminosity upper limit of  $2.6\times10^{30}$ erg s$^{-1}$ \citep{mer11b}. 
Also a  search for radio emission from  this system gave a  negative result \citep{coe11}.

 \begin{figure}
\begin{center}
\includegraphics[width=8.5cm,angle=0]{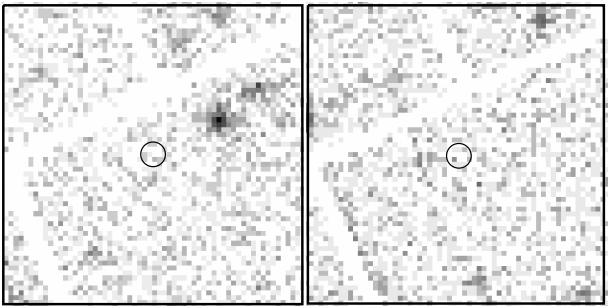}
\caption{Images of the regions of \cd\ (left panel) and \pg\ (right panel) in the 0.2--2 keV energy range obtained with the EPIC pn camera.  Both images have a size of 4$\times$4 arcmin$^2$, north to the top, east to the left. The positions of the two sdB binaries are indicated by the circles with radius 10$''$.}
\label{fig_ima}
\end{center}
\end{figure}

\section{Observations and data analysis}

The fields of \cd\ and \pg\  were  observed with  the \xmm\ satellite in August 2013 and  July 2009, respectively. The details of the observations are given in Table \ref{obs-log}. 
We used the data obtained with the EPIC instrument, which consists of one pn \citep{str01} and two  MOS CCD cameras \citep{tur01} operating in the 0.2--12 keV energy range.
In both observations the three cameras were used in full frame mode. The medium thickness optical blocking filter was used in the MOS1 observation of \cd , while all the other data were taken with the thin filter.
The data were processed with Version 12  of the Standard Analysis Software (\textsc{sas}).
   
We checked the data for the presence of high background caused by  soft proton flares, 
finding that only the observation of \pg\ was affected.  We removed the corresponding time intervals,
resulting in the net exposure times given in Table   \ref{obs-log}.
After this standard screening of the data, we accumulated pn and MOS images in different energy bands using only single and adjacent bi-pixel events.  Many X-ray sources were visible in these images,  but none  at the positions of  \cd\ and \pg\ (see Fig. \ref{fig_ima}) . The closest detected sources are at angular separations of $\sim$$1'$  and $\sim$$2'$ from our targets, respectively.

To perform a more   quantitative analysis and to derive luminosity upper limits for our targets, we carried out a source  detection procedure over the whole field of view of the two observations. 
We considered the data of the three cameras simultaneously, in order to maximize the signal-to-noise ratio  of the serendipitous sources and to reach lower flux limits. 
We first performed the source detection in the  0.5--4.5 keV energy range.
For each camera, we used  the cleaned event files to accumulate the field image; then we generated with the \textsc{sas} task \textsc{eexpmap} the corresponding exposure map, to account for spatial quantum efficiency  variations, mirror vignetting, and effective field of view. 
For each camera we run the  \textsc{sas} task \textsc{eboxdetect} in {\em  local mode} to create a preliminary source list. 
Sources were identified by applying the  standard {\em minimum detection likelihood} criterium: for each source we calculated a detection likelihood $L = -\ln P$, where {\em P} is the probability of a spurious detection due to a Poissonian random fluctuation of the
background. 
We considered a threshold value $L_{\mathrm{th}}=7.3$, corresponding to a 3$\sigma$ probability that the detected counts originate from a background  fluctuation. Then, the task \textsc{esplinemap} was  run to remove all the validated sources from the original image and to create a background map by fitting the so called {\em cheesed image} with a cubic spline. Finally, the task \textsc{eboxdetect} was applied again in {\em map} mode on the three images simultaneously, using as a reference the corresponding background maps. 
In this way, the likelihood values from each individual instrument were  converted to  equivalent single-instrument detection likelihoods, and a threshold value of 7.3 was applied to accept or reject a  source.
In order to maximise the sensitivity for particularly soft or hard sources, we repeated the above procedure also in the energy ranges 0.2--0.5 keV, 0.2--2 keV and 2--12 keV. 

No sources were detected at the position of \cd\ and \pg . Therefore  
we used the task \textsc{esensmap} to create the sensitivity maps, which provide   upper limits for the count rates  of the undetected sources as a function of their position in the field of view. 
In this way we obtained 3$\sigma$ upper limits of 0.0016 and 0.0019 counts s$^{-1}$ for the 0.5--4.5 keV  count rate of \cd\ and \pg , respectively. Similar values were obtained for the  upper limits in the other energy bands (see Table \ref{obs-log}).

\begin{table*}
\centering \caption{Log of the \xmm\ observations and count rate upper limits (3$\sigma$).}
\label{obs-log}
\begin{tabular}{@{}llccccccccc}
\hline
Target &    Date     & \multicolumn{2}{c}{Start/end time (UT)} & \multicolumn{3}{c}{Net exposure time (ks)} & \multicolumn{3}{c}{EPIC count rate (counts ks$^{-1}$)} \\
           &     (YYYY-MM-DD)  &\multicolumn{2}{c}{(hh:mm)}                &          pn    &     MOS1  & MOS2  & 0.2--0.5 keV & 0.5--4.5 keV  &  0.2--2 keV   & 2--10 keV\\
\hline
\cd\ &                           2013-08-17  & 04:19  &  18:45  &  45 & 51  & 51  & $<$1.1 & $<$1.6 & $<$1.5  & $<$1.3 \\
\pg\ &                          2009-07-09  & 08:38  &  20:57   &  25 & 36  & 36  & $<$1.4 & $<$1.9 & $<$1.9  & $<$2.0 \\

\hline
\end{tabular}
\end{table*}

\begin{figure}
\begin{center}
\includegraphics[width=6.5cm,angle=90]{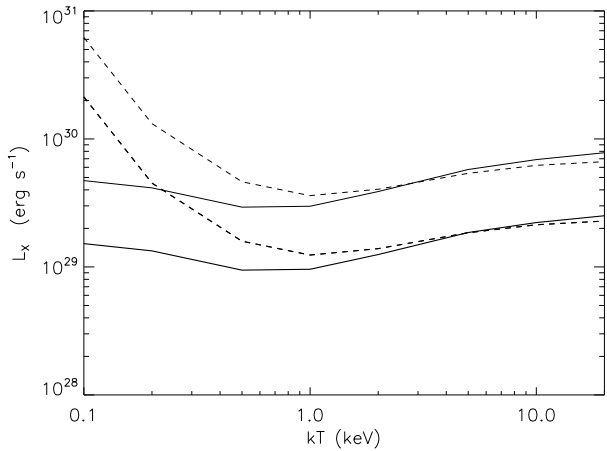}
\caption{Upper limits (3$\sigma$) on the 0.2-10 keV unabsorbed X-ray luminosity of \cd\ (lower lines)  and \pg\ (upper lines) for a thermal bremsstrahlung spectrum of temperature $kT$ and $N_{\rm H}=4\times10^{20}$ cm$^{-2}$. The solid  and dashed lines correspond to the count rate upper limits in the 0.2--2 keV  and 0.5--4.5 keV bands, respectively.}
 \label{fig_ul}
 \end{center}
 \end{figure}

\section{Results}

\subsection{ \cd\  }
 
No X-ray emission from \cd\ was detected in our observation, despite the relatively long exposure time and good sensitivity of the   EPIC instrument. 
To convert the count rate  upper limit to an  X-ray luminosity it is necessary to make some assumptions on the spectral shape of the source and on the interstellar absorption.
\cd\ is at an estimated distance of  364 pc \citep{gei13} and is only slightly reddened,  with $E(B-V)=0.04$ \citep{nem12}. The total column density  in the direction of \cd\ (Galactic coordinates  $l$ = 322$^{\circ}$, $b$ = $+$29$^{\circ}$)  is $N_{\rm H}=4\times10^{20}$ cm$^{-2}$ \citep{kal05}.  In the following  we will  assume this  $N_{\rm H}$ value for the count rate to flux conversions.
 
For a power-law spectrum with photon index $\Gamma$ = 2,  the count rate upper limits  derived above correspond to a source flux of $\sim$$10^{-14}$ erg cm$^{-2}$ s$^{-1}$  (0.2--10 keV, corrected for the absorption). Given the source distance of 364 pc, this implies an X-ray  luminosity  $L_{\mathrm{X}}<1.5\times10^{29}$ erg s$^{-1}$.
A thermal bremsstrahlung spectrum, as typically observed in accreting white dwarfs, leads to similar upper limits on  $L_{\mathrm{X}}$. These  are shown in  Fig. \ref{fig_ul} as a function of the assumed bremsstrahlung temperature $kT$. For temperatures below a few keV the strongest constraints are obtained by the upper limits derived in the soft spectral band image.  
It can be seen that  for temperatures between 0.1 and 20 keV the luminosity is in the range  (1--2)$\times$10$^{29}$ erg s$^{-1}$. We therefore will use  $L_{\mathrm{X}}=10^{29}$ erg s$^{-1}$ as a reference normalization in the following considerations.

To derive some constraints on the mass loss rate $\mdot_{\mathrm{W}}$  from \cd\ based on the  $L_{\mathrm{X}}$ upper limits, it is necessary to consider the mechanism of mass transfer and accretion possibly at work in this binary system.  Even if the subdwarf radius is close to that of its  Roche-lobe ($\sim$83\%), we will for simplicity consider only accretion from the stellar wind and adopt the Bondi-Hoyle formalism.   
In this case the accretion luminosity for a WD of mass $M_{\mathrm{WD}}$  and  radius $R_{\mathrm{WD}}$ moving with velocity $v_{\mathrm{rel}}$ with respect to the wind is given  by:

$$ L_{\mathrm{X}} = \frac{  G M_{\mathrm{WD}} }{ R_{\mathrm{WD}}  }  \mdot_{\mathrm{W}} \bigg( \frac{R_{\mathrm{a}}}{2 a} \bigg)^2 $$
   
\medskip

\noindent
where $a$ is the orbital separation and $R_{\mathrm{a}} =2GM_{\mathrm{WD}}/v_{\mathrm{rel}}^2$ is the accretion radius.
Although this is a good approximation in high mass X-ray binaries, where  $R_{\mathrm{a}} \ll a$, we caution  that it should be considered only as a rough estimate  due to the small orbital separation of this binary.
The X-ray luminosity depends on the wind velocity $v_{\mathrm{W}}$, since $v_{\mathrm{rel}}^2 = v_{\mathrm{o}}^2 + v_{\mathrm{W}}^2$,   where  $v_{\mathrm{o}}\sim630$  km s$^{-1}$  is the WD orbital velocity.

At large distances from the star, the velocity of a radiatively driven wind tends to a maximum value $v_{\infty}$ of the order of the escape velocity.
However,  the wind velocity at the position of the WD is most likely smaller than $v_{\infty}$. For example, a  typical wind velocity law  $v_{\mathrm{W}}(R)=v_{\infty}(1-R_{\mathrm{sdB}}/R)^{\beta}$ with $\beta=1.5$   gives  $v_{\mathrm{W}}/v_{\infty}\sim0.6$ at $r=a$.
Therefore we assume $v_{\mathrm{rel}}=1000$ km s$^{-1}$ and,  for  M$_{\mathrm{WD}}=0.74$ $\msun$ and  R$_{\mathrm{WD}}=0.01$ $\rsun$,   we derive the  relation   

$$ \mdot_{\mathrm{W}} = 2\times10^{-13} \bigg(\frac{L_{\mathrm{X}}} {10^{29} ~ {\rm erg~s^{-1} }}\bigg)  ~~~   {\rm  \msun ~ yr^{-1} } $$ 
                            
\noindent
which can be used to convert the luminosity upper limit derived with \xmm\ into a constraint on  the wind  mass loss rate of \cd  .

\subsection{ \pg\ }

The count rate upper limits derived for \pg\ from our  \xmm\ observation are only slightly higher than those of \cd\ (see Table \ref{obs-log}). Also for this source we assumed an interstellar absorption of   $N_{\rm H}=4\times10^{20}$ cm$^{-2}$, consistent with the total Galactic column density in its direction  \citep{kal05}. Due to the greater distance of \pg , its flux upper limits correspond to X-ray luminosity values about three times larger than those of \cd\  (see Fig. \ref{fig_ul}), i.e. $\sim5\times10^{29}$  erg s$^{-1}$.
 
To constrain $\mdot_{\mathrm{W}}$, we assume the mass lower limit of 6 $\msun$ for the black hole companion of \pg ,  a wind velocity of 1000 km s$^{-1}$ to compute the accretion radius, and an efficiency $\eta$ for the conversion of accretion power to X-ray luminosity.  In this way  we find   $\mdot_{\mathrm{W}}<10^{-13}$ ($L_{\mathrm{X}}/5\times10^{29}$  erg s$^{-1}$)  (0.01/$\eta$)    $\msun$  yr$^{-1}$.

\section{Discussion and conclusions}

Exploiting  the high sensitivity of the \xmm\ EPIC instrument we have carried out  the first deep X-ray observations of two sdB binaries containing compact companions and obtained upper limits on their luminosity much lower than those previously available.
In fact, \cd\ was never pointed before with X-ray satellites and it was not detected in the ROSAT All Sky Survey (0.1--2.4 keV). 
The ROSAT survey data, with  an exposure of only 200 s at its position, give an upper limit of several $10^{31}$ erg   s$^{-1}$. 
A short \sw\ observation of \pg\ \citep{mer11b}   resulted in  an upper limit of $\sim$$3\times10^{30}$  erg s$^{-1}$  (0.3--10 keV),  for a power law with photon index $\Gamma=2$ (about ten times higher than that obtained here for the same spectrum). 
 
The new upper limits on the X-ray luminosity of  \cd\ and \pg , of the order of a few $10^{29}$ erg  s$^{-1}$ for most spectral assumptions
for the count rate to flux conversion,
are the most constraining ever derived for hot subdwarf binaries.
Note that in the case of \cd\ the presence of a WD companion is certain, while the presence of a compact companion in  \pg , as well as in most of the other sdB binaries from which X-ray emission has been searched with \sw\  \citep{mer11b}, relies on the assumption that the sdB star rotates synchronously  with the orbital period (this is required to derive the system inclination which,  coupled to the mass functions, leads to lower limits on the companion mass).   

The lack of detectable X-ray emission in the two observed binaries can be used to derive unprecedented constraints on the wind mass loss from the hot subdwarfs in these systems. 
If the companion to  \pg\ is a black hole we have to face a large  uncertainty on the conclusions that can be drawn, owing to the unknown value of the  efficiency factor $\eta$.  The low X-ray luminosity could indicate a   small mass loss rate from \pg , but it could also be due to a very low efficiency in the conversion of accretion power to X-ray luminosity. This is not unexpected in the case of low rate, spherically symmetric accretion onto a black hole. It is also possible that most of the accretion luminosity is released in a different range, e.g. at much lower energies.
 
In the case of \cd, however, the presence of a  WD of known mass and radius allows us to make more robust considerations. 
The limits on $\mdot_{\mathrm{W}}$ we derived for this sdB  are one of the few observational constraints on the weak winds expected in this class of low mass hot stars. 
It is interesting to make some comparison with the predictions of theoretical models.
\citet{vin02} computed the   mass loss rates for horizontal branch and sdB stars expected from the theory of radiation driven winds.
They derived a scaling relation linking $\mdot_{\mathrm{W}}$ with the star's effective temperature, luminosity, mass and metallicity.  For the parameters of   \cd\ this relation   yields a mass loss rate $\mdot_{\mathrm{W}}$ = $1.2\times10^{-12} Z^{0.97}$ $\msun$~yr$^{-1}$, where $ Z$ is the metallicity. 
\citet{ham10} derived a mass loss vs. luminosity relation for evolved hot, low mass stars, which are more than ten times  as luminous as the sdB stars.
Extrapolation of that relation (Fig. 10 of \citet{ham10}) yields a mass loss rate for \cd\ consistent
with the prediction of \citet{vin02}.
Our derived upper limit on $\mdot_{\mathrm{W}}$, of the order of $3\times10^{-13}$ $\msun$~yr$^{-1}$ for most spectral assumptions, seems difficult to reconcile with the prediction by  \citet{vin02}  if \cd\ has a solar, or higher, metallicity. 
The metal abundance pattern of \cd\  (and \pg ) has not yet been determined. 
In order to make a more meaningful comparison, it would be most important to determine the abundances of heavy elements such as iron, nickel and trans-iron elements, the lines of which dominate the radiation pressure for sdB stars. This determination requires high-resolution UV spectroscopy (see, e.g., \citealt{oto06}).  
We finally note that our upper limit on $\mdot_{\mathrm{W}}$ is in the range of mass loss rates which, according to \citet{ung01}, can explain the He abundances of sdB stars. 

\appendix

\section[]{The serendipitous detection of X-ray emission from the cluster of galaxies \source}

\begin{figure}
\includegraphics[width=0.5\textwidth]{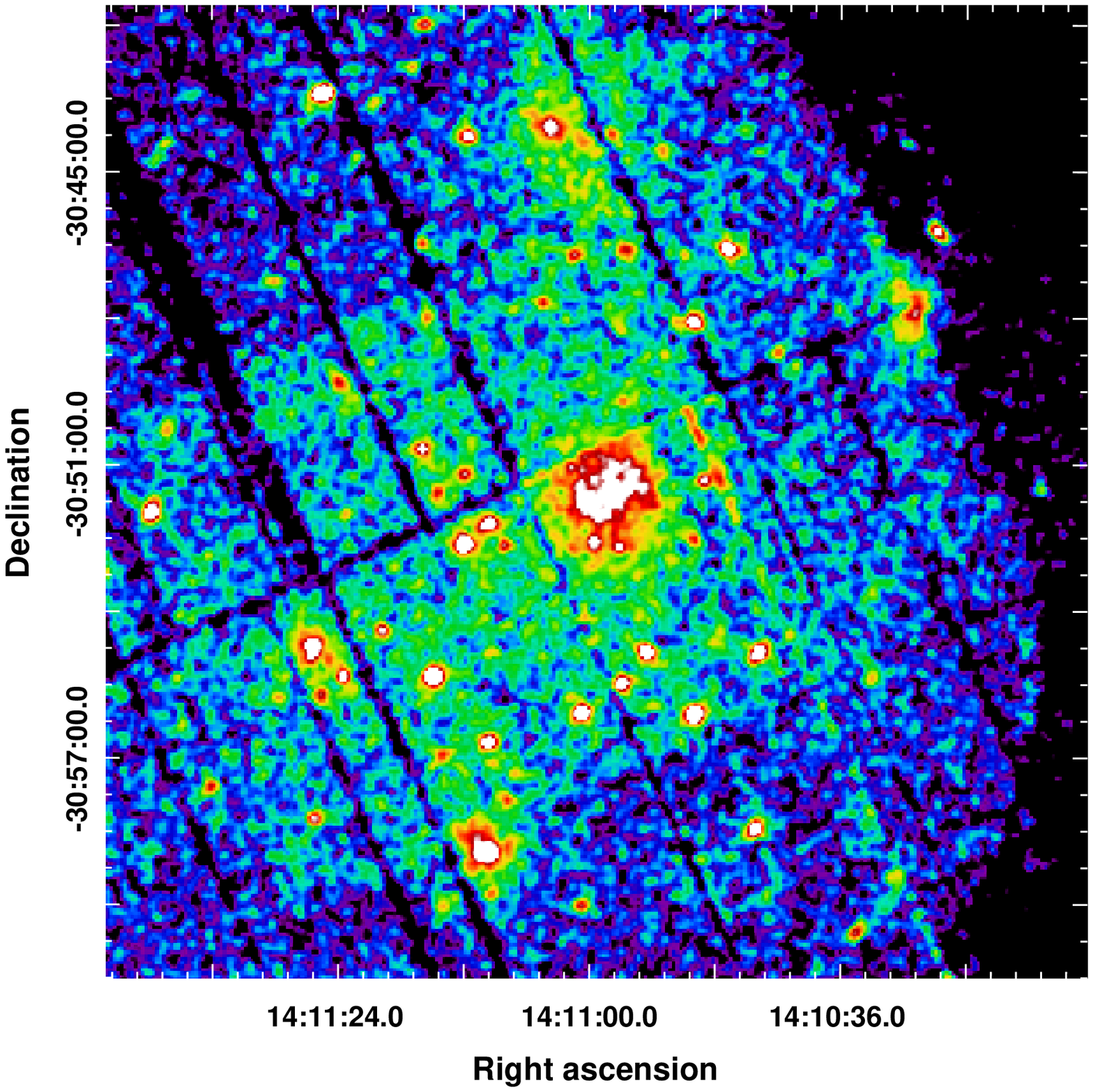}
\caption{EPIC X-ray image in the 0.5--2.0 keV energy band of a 20$'\times20'$ region around the the cluster 
\source\ (see the online journal for a color version of this figure).}
\label{fig.1} 
\end{figure}

An extended source in the field of \cd\ is clearly detected at 4$'$ off-axis (see Fig.\ref{fig.1}).
The most likely interpretation is that of X-ray emission from a cluster of galaxies.
Based on the position of its X-ray centroid  (\mbox{$\alpha_{\rm J2000}=14^{\rm h}10^{\rm m}57\fs7,\, \delta_{\rm J2000}=-30^{\circ}51' 32\farcs3$}), we denote it \source  .  There are no redshifts in \ned\ possibly associated with this
source, except  for the galaxy 2MASX J14104995-3052321, which is at a distance of 2$'$,  therefore outside the visible extent of the diffuse X-ray emission; its redshift of  $z=0.0619$ is too close given also the indication of the redshift obtained from the X-ray data  (see below).
All the errors quoted in this Appendix are at the 68\% confidence limit and all distance-dependent quantities have been computed assuming \ho = 70 \kmsmpc, \omegam = 0.3 and \omegalambda = 0.7.

\begin{figure}
\begin{center}
\includegraphics[width=5.5cm,angle=-90]{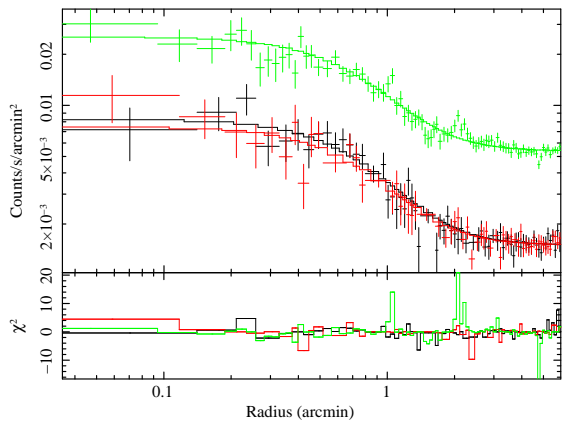}
\vspace{1cm}
\caption{Surface brightness profile of the X-ray emission of \source.
Top panel:  data and best fit beta model. Bottom panel: contribution of each data point to the total $\chi^2$.
Data from MOS1, MOS2 and pn are plotted in black, red and green respectively (see the online journal for a color version of this figure).
}
\label{fig.2} 
 \end{center}
\end{figure}

For each detector we created images in the 0.5--2 keV band with point sources masked using circular regions of 25\arcsec\ radius. Only for three sources embedded in the diffuse emission we used an exclusion radius
of 20\arcsec\ (corresponding to 80\% of the encircled energy fraction at 1.5 keV for a point source at this off-axis angle).
In particular we excluded a hard source (at \mbox{$\alpha_{\rm J2000}=14^{\rm h}10^{\rm m}56^{\rm s},\, \delta_{\rm J2000}=-30^{\circ}51' 03''$}) clearly visible in the 2--5 keV image.
The images have been exposure corrected and a radial surface brightness profile has been extracted from a circular region of 6\arcmin\  radius centered on the cluster position.
We account for the X-ray background in the surface brightness analysis by including a constant-background component. The data were grouped to have at least 20 counts per bin in order to apply the $\chi^{2}$ statistics.
The fitted model is convolved with the \xmm\ point spread function (PSF). The joint best-fit $\beta$-model \citep{Cavaliere.ea:76} has a core radius of $r_{\mathrm{c}} = 56''\pm7''$ and $\beta=0.76\pm0.08$ for a $\chi^{2}$/dof = 287/205 (see Fig.\,\ref{fig.2}). The main contribution to the $\chi^{2}$ comes from the pn and its origin is instrumental (CCD gaps).

\begin{figure}
\includegraphics[width=0.36\textwidth,angle=-90]{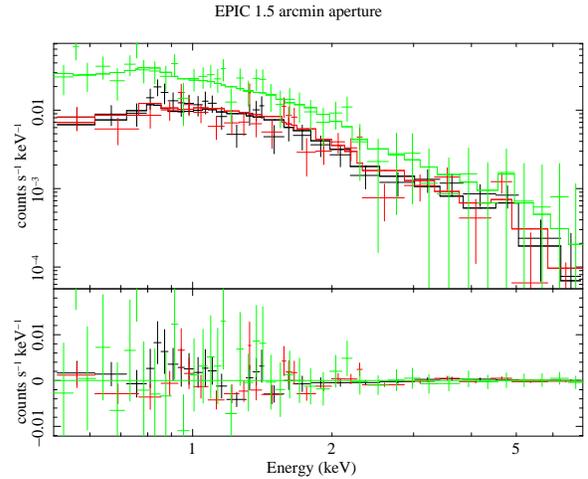} 
\vspace{1cm}
\caption{X-ray spectra of the \source\
taken from a 1.5\arcmin\ radius aperture centered on the centroid of the emission.  The
best fit model and residuals are also shown. Data from MOS1, MOS2 and pn are plotted in black, red and green respectively (see the online journal for a color version of this figure).
}
  \label{fig.3}
\end{figure}

For spectral fitting, we extracted spectra for each detector from a region with 1.5\arcmin\ radius  centered on the emission centroid.
Redistribution matrix files (RMFs) and ancillary response files (ARFs) were generated using the \textsc{sas} tasks \textsc{rmfgen} and \textsc{arfgen} in extended source mode with appropriate flux weighting. The background was estimated locally using spectra extracted from a source free region of the same extent at the same off-axis angle. This is particularly relevant since  the field is in a region (Galactic coordinates $l = 322.43^{\circ}, b=+28.99^{\circ}$) of enhanced Galactic foreground  \citep[see the ROSAT R45 map of][]{Snowden.ea:97}.
The spectra from the three detectors were jointly fitted with an \apec\ thermal plasma \citep{Smith.ea:01} with Galactic absorption fixed at    $N_{\rm H}=4\times10^{20}$ cm$^{-2}$ \citep{kal05}. 
The spectral fitting was performed in the 0.5--7 keV band using the C-statistic and quoted metallicities are relative to the abundances of \citet{Anders.ea:89}. 
The spectra are shown in Fig.\ref{fig.3}: we fixed the metallicity at $Z=0.3$ \solar\ and the best fit parameters are $kT = 4.1\pm0.3$ keV  and $z=0.38\pm0.01$ for a C-stat/dof = 766/716. The X-ray determinations of the redshift  have been proven to be sufficiently reliable when compared with optical redshifts \citep[e.g.,][]{Gastaldello.ea:07,Panessa.ea:09}:  in the following we will therefore assume $z=0.38$ as the redshift of the source, for which 1\arcmin\ corresponds to 311 kpc. 
We obtained a flux of $(8.94\pm0.28)\times10^{-14}$ \fxunits\ in the 0.5--2 keV band in the 1.5\arcmin\ aperture, increased by 12\% due to the area lost for the point source exclusion.   
The source photons correspond to about 50\% of the total events ($\sim$1500 counts in each MOS and 3400 in the pn).

Under the assumption of isothermality  we can calculate the total mass profile using the best-fit $\beta$-model ($r_{\mathrm{c}} = 290\pm36$ kpc), for which the gas density and total mass profiles can be expressed by simple analytical formulae \citep[e.g.][]{Ettori:00}.
We evaluated $r_{500}$ as the radius at which the density is 500 times  the critical density and the virial radius as the radius at which the density corresponds to $\Delta_{\rm{vir}}$, as obtained by \citet{Bryan.ea:98}\footnote{$\Delta_{\mathrm{vir}}=18\pi^2+82x-39x^2$ where $x=\Omega(z)-1$, $\Omega(z)=\Omega_{\mathrm{m}}(1+z)^3/E(z)^2$ 
and $E(z)=\left[\Omega_{\mathrm{m}}(1+z)^3+\Omega_{\Lambda}\right]^{1/2}$.}
for the concordance cosmological model used in this paper.
To evaluate the errors on the estimated quantities we used the procedure of repeating the measurements for 10000 random selections drawn from Gaussian distributions for the temperature and parameters of the surface brightness profile.
For $\Delta=500$ we obtained, $M_{500} = (2.67\pm0.59) \times 10^{14}$  $\msun$ within
$r_{500} = 857\pm63$ kpc; the virial mass is, $M_{\rm{vir}}= (5.90\pm1.18)\times10^{14}$ $\msun$ ,
within the virial radius $r_{\rm{vir}} = 1747\pm116$ kpc.
For \source, the aperture of 467 kpc used for spectroscopy encloses 76\% of the flux within $r_{500}$.
The derived bolometric luminosity within $r_{500}$ is $L_{500} = (1.19\pm0.11) \times 10^{44}$\lxunits.
Given also the limited radial range in which the luminosity and, in particular, the temperature
have been obtained we can only estimate that the self-similar scaled value of $E(z)^{-1}\,L_{500}$ for \source\ is in agreement  with the local \lx--\tx\ relations \citep{Markevitch:98,Arnaud.ea:99,Pratt.ea:09}, with an indication of being slightly underluminous, which might be consistent with recent claims of a negative evolution of the \lx--\tx\ relation  \citep{Reichert.ea:11}.

\section*{Acknowledgments}

We acknowledge the financial support by the Italian Space Agency through
the ASI/INAF agreement  I/032/10/0 for the XMM-Newton operations.
This research is based on data of XMM-Newton, an ESA science mission with instruments and contributions directly funded by ESA Member States and NASA.
 
\bibliographystyle{mn2e}
\bibliography{biblio_HD}

\bsp

\label{lastpage}

\end{document}